\documentclass[aps,twocolumn,aps,amssymb,showpacs,superscriptaddress]{revtex4}
\usepackage{graphicx}
\usepackage{color}
\usepackage{amsmath,amsfonts,amsthm}

\begin{document}

%\title{Giant quantum Goos-H\"{a}nchen effect at normal-superconductor interface}
%\title{Sticky {\color{blue} Goos-H\"{a}nchen effect at} Normal/Superconductor Interface}
\title{Sticky Normal/Superconductor Interface}
\preprint{prepared for PRL}

\author{Soo-Young Lee}
%\email{sooyoung@pks.mpg.de}
\affiliation{Max Planck Institute for the Physics of Complex Systems, N\"{o}thnitzer Str. 38, D01187 Dresden, Germany}
\date{\today}
\author{Arseni Goussev}
\affiliation{Max Planck Institute for the Physics of Complex Systems, N\"{o}thnitzer Str. 38, D01187 Dresden, Germany}
\affiliation{Department of Mathematics and Information Sciences, Northumbria
University, Newcastle Upon Tyne,\\NE1 8ST, United Kingdom}
\author{Orestis Georgiou}
\affiliation{Max Planck Institute for the Physics of Complex Systems, N\"{o}thnitzer Str. 38, D01187 Dresden, Germany}
\author{Goran Gligori\'{c}}
\affiliation{Vinca Institute of Nuclear Sciences, University of
Belgrade, P. O. Box 522, 11001 Belgrade, Serbia}
\author{Achilleas Lazarides}
\email{acl@pks.mpg.de}

\affiliation{Max Planck Institute for the Physics of Complex Systems, N\"{o}thnitzer Str. 38, D01187 Dresden, Germany}
\date{\today}

\begin{abstract}
We study the quantum Goos-H\"{a}nchen(GH) effect for wave-packet dynamics at
a normal/superconductor (NS) interface.
We find that the effect is amplified by a factor $(E_F/\Delta)$,
with $E_F$ the Fermi energy and $\Delta$ the gap. Interestingly, the GH effect
appears only as a time delay $\delta t$ without any lateral shift,
and the corresponding delay length is  about $(E_F/\Delta)\lambda_F$,
with $\lambda_F$ the Fermi wavelength. This makes the NS interface
{\it ``sticky''}  when $\Delta \ll E_F$, since typically GH effects
are of wavelength order.
This ``sticky'' behavior can be further enhanced by a resonance mode in NSNS interface.
Finally, for a large $\Delta$, the resonance-mode effect makes
a transition from Andreev to the specular electron reflection
as the width of the sandwiched superconductor is reduced.
\end{abstract}

\pacs{03.65.Vf, 74.45.+c, 73.40.-c}

\maketitle

%{\color{red} Some introduction, Review on GH shift in Optics and Condensed matter}

%\section{Introduction}
Interference is an important phenomenon in both quantum and wave mechanics, and
 is often responsible for intriguing effects not understandable
from a classical mechanics and ray dynamics.
The Goos-H\"{a}nchen (GH) effect \cite{GH47},
first predicted by Newton, is one example:
When an optical beam is totally reflected at a dielectric interface,
a lateral shift along the interface is induced.
The GH shift has been studied in various quantum and wave systems, amongst which
graphene \cite{Beenakker09,Sharma11}, optical beams \cite{Loffler12}, dielectric microcavities
\cite{Lee05,Schomerus06,Unter08} and neutrons \cite{Haan10}.

%===========================================================
%\section{Goos-H\"anchen shift}
The GH shift in optics can be understood by the dependence
of the phase loss occuring at total internal reflection
\cite{Hawkes95}  on the angle of incidence.
An analogous shift is observed in the quantum mechanical step potential problem:
In the one dimensional case, if the energy $E$ of an incident particle is less than the
potential height $V$, the particle is totally reflected with reflection
coefficient is $R=e^{i\varphi}$ (with real $\varphi$). The phase $\varphi$ increases
monotonically from $-\pi$ to zero as the energy $E$ varies from zero to $V$.
If we consider an incident Gaussian wave packet with velocity $v_0=\hbar k_0/m$,
\begin{equation}
\psi_I (x,t)=\int dk \, g(k) \, e^{ikx}e^{-iEt/\hbar},
\label{Rwfstep}
\end{equation}
where $g(k)=e^{-(k-k_0)^2/\sigma^2}$, $\sigma$ is the width of the Gaussian window, and
$E=(\hbar k)^2/2m$,
the wave packet reflected off the step potential at $x=0$ becomes
\begin{equation}
\psi_R (x,t)=\int dk \, g(k) \, R \, e^{-ikx}e^{-iEt/\hbar}.
\label{Rwfstep}
\end{equation}
The stationary phase approximation gives a time-delayed motion of the centre of the wave packet,
$x=-v_0(t-\delta t)$, with the time delay given by
\begin{equation}
\delta t=\hbar \frac{d\varphi}{dE}.
\end{equation}
As a result, the delay length $l_D \equiv v_0 \delta t$
is $l_D \sim \lambda_0$ with $\lambda_0=2\pi/k_0$ the central wavelength of the
wave packet.
 For a wave packet propagating in two dimensions with $y$-direction velocity $v_{0y}$,
the time delay results in a lateral shift of $l_{GH} =v_{0y}\delta t$,
analogously to the optical GH shift. Note that the GH shift here is of
the order of a wavelength.
This $\delta t$ is similar to the Wigner delay time in scattering problems \cite{Wigner55}.
The phase change $\varphi$ is also related to the non-integer Maslov index
for the quantization condition \cite{Friedrich96}.

%=======================================================================
%\section{Andreev reflection}
On the other hand, the behavior of an electron or a hole at normal/superconductor
(NS) interface has been studied extensively. For a small superconducting pairing gap $\Delta$
($\sim 10^{-4} E_F$) an incident electron from the normal side is retro-reflected
as a hole when the excitation energy is less than the pair gap $\Delta$ of the superconductor;
this is called Andreev reflection \cite{Andreev64}, and has been
studied in systems as diverse as graphene~\cite{Beenakker06} and
bosonic condensates~\cite{ZapataSols}. It has been directly experimentally observed~\cite{experimentsSS}.
In addition, the advance of laser cooling techniques
for atoms allows experimental realization of a similar interface with a large $\Delta$ comparable to
the Fermi energy $E_F$, where specular reflection of particles (in addition to Andreev reflection of holes)
occurs with appreciable amplitude \cite{Achilleas07,experimentsKetterle,experimentsHulet}. In spite of much attention to the electronic
properties at the NS interface, the phase change under reflection has not been
explored so far. It is therefore desirable to understand the influence of the phase
change on reflection off the NS interface and the associated GH effect.

%=============================================================
%\section{Statement of problem}
In this Letter, we study reflection off NS and NSNS interfaces and find that
the GH effect is amplified by a factor $(E_F/\Delta)$ compared to that of other interfaces.
This is because at the NS interface the incident electron can be
reflected within the energy scale of $\Delta$, while the wavelength is
determined by the whole energy scale of $\sim E_F$. Moreover, the GH effect
appears as a time delay without the lateral shift which has been
the hallmark of the GH effect in other interfaces. The corresponding
delay length is then given by $l_D \sim (E_F/\Delta)\lambda_F$,
where $\lambda_F$ is the Fermi wavelength,
implying that the NS interface is very {\it ``sticky''} in comparison
to typical wavelength-order delay lengths in optical and condensed-matter interfaces.
We also show that the time delay can be increased
by a resonance mode in NSNS interface. Finally, for a large gap $\Delta$  ($\sim E_F$)
the resonant reflection exhibits a transition from Andreev retroreflection to
specular electron reflection as reducing the width of sandwiched superconductor. 

\begin{figure}
\includegraphics[width=3.0in]{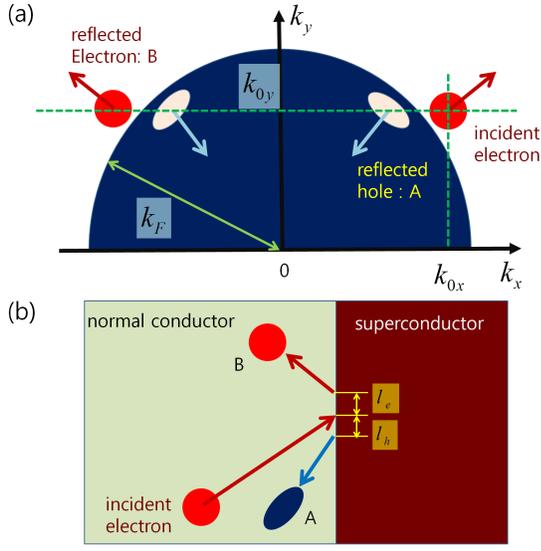}
\caption{(color online) (a) Schematic diagram for electron and hole excitations
near the Fermi surface in $(k_x,k_y)$ plane. The arrows indicate the direction of
group velocity of the corresponding wave packets.
(b) The real-space illustration of the excitations in (a).
An incident electron can be bounced into a hole (A) and an electron (B).
$l_e$ and $l_h$ are GH shifts expected from a naive extension of the step-potential case,
but they turn out to be zero in this NS interface (see Fig. \ref{fig2} and \ref{fig4}).
}
\label{fig1}
\end{figure}
%\section{Outline of paper}
In what follows, we describe our theoretical setup, then begin our
study with the
$\Delta/E_{F}\ll 1$ for both NS and NSNS interfaces. We then turn to
the general $\Delta/E_{F}$ case, for which we discuss the crossover
between Andreev and specular reflection before drawing our
conclusions.

%\section{BdG equations}
%===============================================================
In a superconductor, electrons and holes are coupled to each other with
a coupling strength given by the pair gap $\Delta$; this situation is described using the
Bogoliubov-de Gennes (BdG) equations for a two-component
wave function $\Psi$ \cite{Gennes99,BTK}:
\begin{equation}
\begin{pmatrix}
   H     & \Delta   \\
  \Delta &   -H
\end{pmatrix}
\Psi = \varepsilon \, \Psi ,
\end{equation}
with
\begin{equation}
H=-\frac{\hbar^2}{2m} \nabla^2 -\mu, \quad \Psi=\binom{u}{v}
\end{equation}
where $\varepsilon$ is the energy measured from the chemical potential $\mu$
($\mu=E_F$ in our case) while
$u$ and $v$ are the electron and hole wave functions, respectively.
The pair gap $\Delta$ is zero in a normal conductor, i.e., the electron
and hole are independent of each other. We also assume that $\Delta$ is a real
number, that is, we consider an $s$-wave superconductor and a static situation.

%======================================================

%\section{NS, plane waves, $\Delta\ll E_{f}$}

We begin by focussing on the case of an incident electron with an energy $\varepsilon$ ( $< \Delta$)
as schematically illustrated in Fig. \ref{fig1}.  From the BdG equation, the two independent plane-wave solutions \cite{Andreev64,McMillan68}
with the same $y$-dependence $e^{ik_y y}$ are
\begin{equation}
\Psi_e = \binom{1}{\frac{\varepsilon -\Omega}{\Delta}} e^{i(k_e x+k_y y)},
\quad
\Psi_h = \binom{\frac{\varepsilon -\Omega}{\Delta}}{1} e^{i(k_h x+k_y y)},
\label{wfeh}
\end{equation}
where $\Omega=\sqrt{\varepsilon^2-\Delta^2}$ and
\begin{equation}
k_e=\sqrt{k_{Fx}^2 +\frac{2m}{\hbar^2}\Omega},
\quad
k_h=\sqrt{k_{Fx}^2 -\frac{2m}{\hbar^2}\Omega},
\label{keh}
\end{equation}
where $k_{Fx}=\sqrt{\frac{2m}{\hbar^2}E_F-k_y^2}$ and it becomes
Fermi wavenumber $k_F$ when $k_y=0$. Note that in the normal conductor
with zero $\Delta$, the solutions become independent electron ($u$) and hole ($v$)
wavefunctions since $(\varepsilon-\Omega)/\Delta=0$ in Eq.(\ref{wfeh}), and their wavenumbers,
$k_e^N$ and $k_h^N$, are given by Eq.(\ref{keh}) with $\Omega=\varepsilon$.

We obtain the reflection coefficients for the reflected electron
and hole plane waves by matching wavefunctions at the interface ($x=0$).
The wavefunction in the normal part is
\begin{equation}
\Psi_N = \binom{e^{ik_e^N x}+Be^{-ik_e^N x}}{Ae^{ik_h^N x}} e^{ik_y y},
\end{equation}
while the wave function in the superconductor is $\Psi_S$ with decaying $u$ and $v$
for our case of $\varepsilon < \Delta$.  From $\Psi_N (0)=\Psi_S (0)$ and
$\Psi'_N (0)=\Psi'_S (0)$, we can obtain the reflection coefficients $A$ (hole)
and $B$ (electron). 

%\section{NS, wavepacket, $\Delta\ll E_{f}$}
In order to study a dynamical situation, we now switch to
an incident wave packet rather than a plane wave.
The incident electron with an average wavevector $\vec{k}_0=(k_{0x},k_{0y})$
can be described as a Gaussian wave packet,
\begin{equation}
u_I (\vec{x},t) = \int dk_x dk_y \, {\cal G}(\vec{k}) \, e^{i(k_x x+k_y y-\varepsilon t/\hbar)}
\label{uin}
\end{equation}
where ${\cal G}(\vec{k})=e^{-|\vec{k}-\vec{k}_0|^2/\sigma^2}$
and $\varepsilon=(\hbar^2|\vec{k}|^2/2m) -E_F$. Note that at $t=0$ the wave packet
hits the interface located at $x=0$.
The reflected hole is then represented by
\begin{equation}
v_R (\vec{x},t) =  \int dk_x dk_y \, {\cal G} (\vec{k}) \, A \, e^{i(k_{h}^N x+k_y y-\varepsilon t/\hbar)},
\label{holewp}
\end{equation}
and the wave packet for the reflected electron can be obtained by the replacements
$A\rightarrow B$ and $k_{h}^N \rightarrow -k_x$ in the above equation (see Fig. \ref{fig1} (a)).
For the normal incidence case $k_{0y}=0$, it is easy to get time delay from the
stationary phase point: The peak position of the reflected hole is given by
\begin{equation}
x=\frac{v_{0x}}{d k_h^N/dk_x} (t-\delta t^A_x), \quad \delta t^A_x =\hbar \frac{d\varphi_A}{d\varepsilon},
\end{equation}
where $\varphi_A$ is the phase of $A$. Similarly, for the reflected electron we find
$\delta t^B_x =\hbar (d\varphi_B/d\varepsilon)$.

\begin{figure}
\includegraphics[width=2.8in]{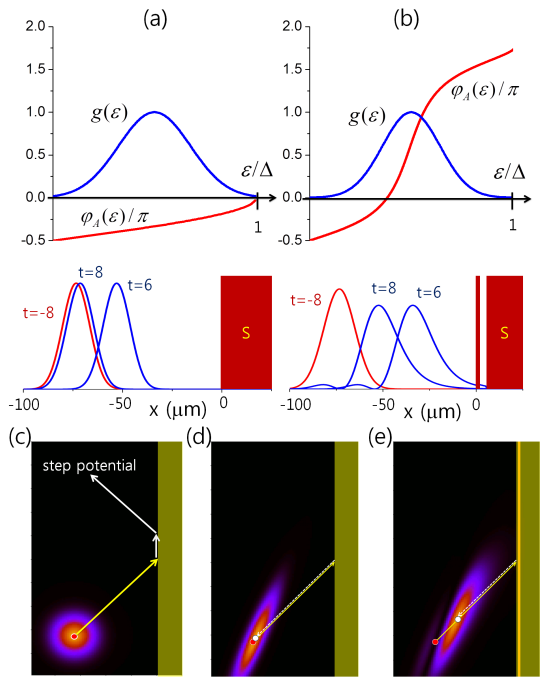}

\vspace{-0.5cm}

\caption{(color online) Wave-packet dynamics when $\Delta=10^{-4} E_F$.
(a) The phase $\varphi_A$ (red line) and the Gaussian
window function $g (\varepsilon)$ (blue line) used.
In the lower pannel, electron wave packets (red line) at $t=-8 $
and reflected hole wave packets (blue lines) at $t=6, 8$ are shown.
(b) The same as (a) for the resonant reflection in the NSNS interface
with $(d,L)=(1,4)$.
(c,d,e) Wave-packet dynamics with 45 degree incident angle.
The scale of the normal conductor shown above is $(100 \times 200)\, \mu m^2$.
(c) The incident-electron wave packet at $t=-10$. The yellow arrow denotes the incident
path. The white arrows show schematically a typical GH shift of the step-potential case.
(d) Andreev reflection. The reflected-hole wave packet at $t=10$ is shown.
(e) The same as (d) in the NSNS interface with $(d,L)=(0.8,2.8)$.
The delay length corresponds to the distance between initial (red dot) and final position
(white dot).
Here, the units of length and $t$ is $\mu m$ and $6.58 \times
10^{-12}s$, respectively.
(see supplemental material for wave-packet animations)
}
\label{fig2}
\end{figure}

%======================================================================
We now concentrate on the case of relevance to solid-state
and liquid helium experiments~\cite{experimentsSS}, $\Delta\ll
E_{F} $, specifically with $\Delta/E_F=10^{-4}$ which is the typical value in solid-state superconductors.
We set the Fermi energy as $E_F=5.5 \, eV$ throughout
this Letter. This case corresponds to Andreev reflection \cite{Andreev64},
so that only the  retro-reflected hole has appreciable amplitude ($|A|\simeq 1$ and $|B| \simeq 0$). The phase
$\varphi_A$ is drawn in Fig. \ref{fig2} (a) and it is well approximated by
$\varphi_A=\arccos(\varepsilon/\Delta)$ when $B=0$. This phase variation enables us to calculate
the delay time and  length, defined by $l_D \equiv v_F \delta t_x^{A}$,
$v_F$ is the Fermi velocity. We emphasize that the delay length is amplified by a factor
$(E_F/\Delta)$, i.e.,
\begin{equation}
l_D \simeq \left(\frac{E_F}{\Delta}\right) \lambda_F,
\end{equation} 
which is a distinctive feature of the NS interface, noting that typical shifts
in other interfaces are wavelength-order. Thus the NS interface can be characterized as {\it ``sticky''} due to the long-time stay at the interface.

\begin{figure}
\includegraphics[width=2.8 in]{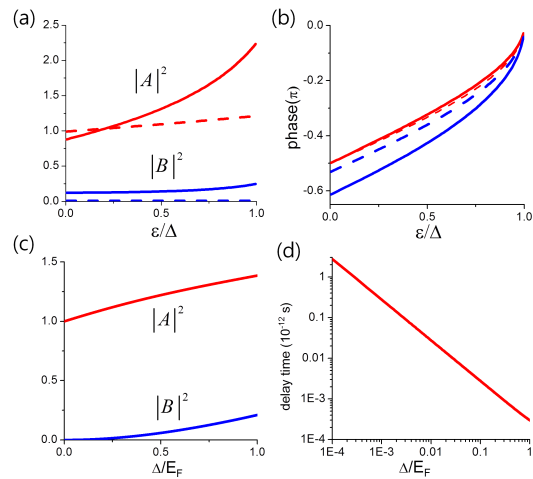}
\caption{(color online) (a) $|A|^2$ (red) and $|B|^2$ (blue) are shown
as a function of $\varepsilon$,
when $\Delta=0.2E_F$ (dashed line) and $\Delta=0.8E_F$ (solid line).
(b) The corresponding phases, $\varphi_A$ (red) and $\varphi_B$ (blue).
(c) Change of $|A|^2$ (red) and $|B|^2$ (blue) evaluated at $\varepsilon =0.5 \Delta$
with increasing $\Delta$. (d) The corresponding delay time $\delta t^A$ as a function of $\Delta$.
}
\label{fig3}
\end{figure}

%===========================================================================
\begin{figure}
\includegraphics[width=2.8 in]{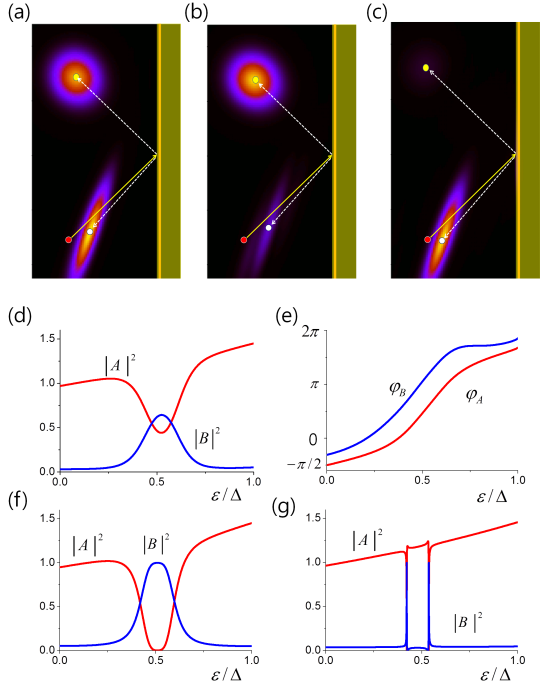}
\caption{(color online) Various resonant reflections when $\Delta=0.2E_F$.
(a) $(d,L)=(3 ,13)$. (b) $(d,L)=(4.7,13 )$.  (c) $(d,L)=(20,13 )$ in
the unit of $\AA$.
The red dots denote the initial position and white and yellow
dots correspond to the final positions of hole and electron, respectively.
Note that the incident-electron velocity is not parallel to the reflected-hole
velocity in this large $\Delta$ case, because both are parallel to radial lines
of Fermi circle as shown in Fig. \ref{fig1} (a).
The scale of the normal conductor shown above is $(0.12 \mu m \times 0.24 \mu m)$.
(d),(f),(g) $|A|^2$ and $|B|^2$ as a function of $\varepsilon/\Delta$
for the cases of (a),(b),(c), respectively.
(e) The phases, $\varphi_A$ and $\varphi_B$ for the $(d,L)=(3,13)$
case (see supplemental
material for wave-packet animations).
}
\label{fig4}
\end{figure}

%=================================================================
%\section{NSNS, plane waves, $\Delta\ll E_f$}
Since the delay time is proportional to the slope of the phase change,
 $d\varphi_{A(B)}/d\varepsilon$, it can be extended by increasing the slope.
This may be done with the help of a resonance mode introduced by changing
the NS interface into a NSNS structure. We denote the widths of first superconductor barrier and
next normal conductor as $d$ and $L$, respectively. 
The phase $\varphi_A (\varepsilon)$ for plane waves in the NSNS case is shown in
Fig. \ref{fig2} (b): the slope becomes steeper as $d$ increases, eventually
approaching an abrupt $2\pi$ phase jump at the resonance point.

%===================================================================
%\section{NSNS, wavepacket, $\Delta\ll E_f$}
Now consider a Gaussian wavepacket,
beginning with the normally incident case, $\vec{k}_0=(k_{0x},0)$.
For the window function ${\cal G} (\vec{k})$ in Eq. (\ref{uin}),
we take $k_{0x}=\frac{k_F+k_M}{2}$, where $k_M=\sqrt{k^2_F +\frac{2m}{\hbar^2} \Delta}$,
and $\sigma=\frac{k_M-k_F}{n}$ , $n=4$ and 5 for the NS and NSNS interface.
The $k_x$ dependence of the window function is given by $g(k_x)={\cal G}(k_x,0)$, and
it corresponds to $g(\varepsilon)=e^{-(\varepsilon-\varepsilon_0)^2/\sigma_\varepsilon^2}$, $\varepsilon_0=\Delta/2$
and $\sigma_\varepsilon=\Delta/n$, which are shown as blue line in the upper pannels in \ref{fig2} (a),(b).
The wave-packet movements are shown in the lower pannels.
The red line and blue line describe
 the incident-electron and reflected-hole wave packets, respectively.
For the NS interface in Fig. \ref{fig2} (a), we can see some distance between wave packets of
electron at $t=-8$ and hole at $t=8$, correspondig to the delay length,
$l_D=v_F \delta t$ (note $|v_e| \simeq |v_h| \simeq v_F$ in this case, $v_e$ and $v_h$ are
electron and hole group velocities).
For the resonant reflection in Fig. \ref{fig2} (b),
a longer time delay is expected from the steeper slope of $\varphi_A (\varepsilon)$,
and the extended delay length is clearly shown in the lower panel of Fig. \ref{fig2} (b).

%========================================================================

Now consider a finite incidence angle, i.e., $k_{0y} \ne 0$; here one might
expect a GH shift for the reflected hole like the $l_h$ in Fig. \ref{fig1} (b),
based on the separablility of $x$ and $y$ coordinates like for a diagonal step potential.
However, this is not the case.
The wave-packet dynamics are illustrated in Fig. \ref{fig2} (c)-(e).
Figure \ref{fig2} (c) shows the incident wave packet with 45 degree incident angle.
The retro-reflected hole packets are shown in Fig. \ref{fig2} for the NS interface (d) and
for the NSNS interface (e), for two times symmetrically before and after the time of impact on the interface. From the distance $l_D$ between initial and final position of the wave packet it is clear that there is a time delay, but no GH shift at all.
This can be understood by the dependence of the energy $\varepsilon$ on $k_y$, i.e.,
$\varepsilon=\frac{\hbar^2}{2m}(k_x^2+k_y^2) -E_F$.
This gives rise to a time delay $\delta t^A_y$ in the $y$-direction too, and this time delay
is the same as $\delta t^A_x$. Thus we see only time delay without GH shift in Fig. \ref{fig2} (d),(e). The deformed shape of the hole wave packet comes from the relation between $k_x$ and $k^N_h$. In other words, the symmetric Gaussian window ${\cal G}(k_x,k_y)$ becomes
(approximately) an asymmetric Gaussian windows in $(k^N_h,k_y)$ plane, as depicted in Fig. \ref{fig1} (a). 
%===============================================================
%\section{NS, plane waves, $\Delta\approx E_{f}$}

We now turn to a discussion of the more general case where $\Delta/E_{F}$ is
not small. As $\Delta/E_F$ increases the nature of the reflected
object changes from purely Andreev- to mixed Andreev- and 
specularly-reflected~\cite{Achilleas07}.  Figure \ref{fig3} shows how the 
coefficients $A$, $B$ and the delay time $\delta t$ change
as $\Delta$ increases.  $|A|^2$ and $|B|^2$ and the phases $\varphi_A$ and
$\varphi_B$ are shown, as a function of $\varepsilon$, in Fig. \ref{fig3} (a),(b),
where the dashed lines are for $\Delta=0.2E_F$ case and
the solid lines are for $\Delta=0.8E_F$ case.
When $\Delta=0.2 E_F$, the coefficents indicate almost pure Andreev reflection
($|A|=1$ and $|B|=0$). At $\Delta=0.8 E_F$, $|B|^2$
becomes appreciable so specular reflection becomes important. 
% The two coefficents $A$ and $B$ are related by current conservation
% $(k^N_h/k^N_e)|A|^2 + |B|^2=1$ \cite{Achilleas07}. 
Fig. \ref{fig3} (b) implies that  $(d\varphi_{A,B}/d\varepsilon) \sim
1/\Delta$ so that the delay time behaves as $\delta t \sim 1/\Delta$, which is drawn
 in  Fig.~\ref{fig3} (d). In Fig.~\ref{fig3} (c),  $|A|^2$, $|B|^2$ measured at $\varepsilon=0.5\Delta$
are shown as a function of $\Delta$, indicating again the non-negligible
specular electron reflection for a large $\Delta$.

%\section{NSNS, $\Delta\approx E_{f}$}

It is interesting that the resonant reflection in the large $\Delta$ case
exhibits a dramatic change of the reflected object depending on the width of the
superconductor barrier, $d$. As an example, the various reflections, when $\Delta=0.2E_F$,
are shown in Fig. \ref{fig4}. Here we take $L=13 \AA$, which supports a resonance mode
around $\varepsilon=0.5 \Delta$ for the 45 degree incident angle. As shown in Fig. \ref{fig4} (a),
both electron and hole are reflected out at $d=3 \AA$, and then electron-dominant reflection
is observed at $d=4.7 \AA$, contrast to Andreev reflection of the small $\Delta$ case.
For $d=20 \AA$, we see the hole-dominant reflection, back to Andreev reflection.

This rather peculiar behavior can be understood as an effect of the resonance mode.
Figures~\ref{fig4} (d-g) show variation of $|A|^2$ and $|B|^2$ around the resonant point
for the above three different cases. For thin barrier
case, $d=3 \AA$, the resonance mode is quite leaky so that the resonance mode
affects on the coefficients within a rather broad range in $\varepsilon$,
resulting comparable values of  $|A|^2$ and $|B|^2$ at the resonant point (see Fig. \ref{fig4} (d)).
Their delay time has a similar value as expected from the slope of the phase
variation in Fig. \ref{fig4} (e). As increasing the barrier width,
the broad range is getting narrower and reach its maximum, $|A|^2=0$ and
$|B|^2=1$ (see Fig. \ref{fig4} (f)). If the barrier width is getting thicker, the single
peak structure starts to split into a double-peak structure, showing $|A|^2=0$ and
$|B|^2=1$ at each peak positions, and the width of peaks is getting narrower as shown
in Fig. \ref{fig4} (g). This double-peak structure of a resonance mode
is a charactericstic feature of two-component resonance where each component
has its own wavelength \cite{Lee12}.

%========================================================================
%\section{Conclusions}
In conclusion we have studied wave-packet dynamics at NS interface,
and found that the GH effect is amplified by the factor $E_F/\Delta$.
Interestingly, the effect appears only as a large time delay, hence {\it ``sticky''},
without any lateral shift. We also demonstrated that in NSNS interface
the GH effect is even further enhanced by a resonance mode and
for a large $\Delta$ the transition from Andreev to specular electron
reflection occurs when the decay rate of the resonance increases.
Although we have focussed on the incident electron case,  all intriguing features
discussed in this Letter can also be found in the incident hole case.
We believe that the time-delay effect can be confirmed by an
experimental observation, since direct observation of Andreev-reflected electrons
in clean systems (where the motion is practically ballistic) is
possible~\cite{experimentsSS}.

\emph{Acknowledgements}
G.G. acknowledges support from the Ministry of Education and Science of
Serbia (Project III45010); A. L. thanks O. Tieleman for stimulating discussions.

%====================================================================

\end{document}